\documentclass[times, twoside, watermark]{Arxiv-Paper}
\usepackage{blindtext}

\leadauthor{John M. Bass} 

\begin{document}

\title{Exploiting Phase Light Modulators for Low-SWaP Real-time Wavefront Correction at High-Resolution}
\shorttitle{}

\author{John M. Bass}
\author{Jeffrey Y. Chen}
\author{Gregory M. Nero}
\author{Vijay Nafria}
\author{Youngsik Kim}
\author{Ivan B. Djordjevic}
\author{Yushi Kaneda}
\author{Thomas L. Koch}
\author{Muralidhar M. Balaji}
\author{Yuzuru Takashima}
\author{Florian Willomitzer}

\affil{Wyant College of Optical Sciences, University of Arizona, Tuczon, AZ, USA\\
Further author information: (Send correspondence to J.M.B.)\\J.M.B.: E-mail: jmbass@arizona.edu\\  F.W.: E-mail: fwillomitzer@arizona.edu}

\maketitle

\begin{abstract}
Wavefront correction and beam tracking are critical in applications such as long-range imaging through turbulence and free-space optical communication. For instance, adaptive optics systems are employed to correct wavefront distortions caused by atmospheric turbulence and optical misalignments, while beam tracking systems maintain alignment between separate devices in free-space communication scenarios. Current state-of-the-art approaches offer several high-performance solutions, each tailored to specific correction tasks. However, integrating all these functionalities into a single device, e.g., for simultaneous adaptive wavefront correction and tracking, can significantly increase the size, weight, and power consumption (SWaP) of the final system. In this contribution, we demonstrate the use of the Texas Instruments Phase Light Modulator (PLM) as a low-SWaP, chip-scale solution for simultaneous wavefront correction and beam tracking in real-time, featuring over one million actuators. In particular, we will present and discuss multiple algorithms and optimization strategies that we have specifically developed for PLM-based wavefront correction.
\end {abstract}


\section{INTRODUCTION}
\label{sec:intro}  
Free space optical communication (FSoC) technologies currently show promise for the secure establishment of high-data-rate communications beyond 1 gigabit per second, including in signal-contested environments\cite{hawraa_2023_review}. To establish links, however, FSoC systems require precise alignment of the laser beams used to communicate information. These alignment precisions can be microradian-scale, depending on the distance between the transmitter and reciever \cite{harris_2005_alignment}. In general, longer distances require higher-precision alignment.

While beam alignment can be done "by hand" through trial and error for short-range communication links, long-range communication links have many angles that must be checked, making manual alignment infeasible. Thus, it has become common for long-range communication links to use automatic search-and-acquisition systems to perform alignment. These systems rely on fast steering mirrors, optical phased array antennas, or other wavefront modulation devices to rapidly steer a transmitted beam in a desired direction, and most commonly use a method of feedback, such as through a retroreflector or concurrent radio link \cite{kaymak_2018_survey}, to know when the link has been established.

After establishing an alignment, acquisition systems often will occasionally reestablish link alignment to account for slowly-changing sources of beam alignment like thermal expansion of some building materials \cite{hicham_2021_thermal}, or atmospheric ducting \cite{abdullah_2019_chromatic}. Some acquisition systems may additionally perform continuous tracking of the receiver, to allow communication between moving terminals.

FSoC technologies similarly may use continuous wavefront correction, using adaptive optics (AO) techniques \cite{hampson_2021_adaptive,wang_2018_performance}, to minimizes the losses from a communication link through atmospheric turbulence. However, instead of only adjusting the beam tilt, AO systems additionally correct higher-order aberrations such as defocus and astigmatism which may be induced by the atmosphere \cite{wang_2018_performance}. To correct these higher-order aberrations, an AO system first measures the magnitude of each aberration, using a wavefront sensor. Then, these measurements are used to provide feedback to a deformable mirror, which phase conjugates, and thus removes, the aberrations. AO systems have also been widely explored for astronomical imaging, microscopy, and vision science for their capability of improving the resolution of captured images \cite{hampson_2021_adaptive}.

While acquisition, tracking, and AO systems have been highly explored in the literature, each respective system relies on specialty devices to achieve maximum performance. Acquisition and tracking systems tend to rely on either fast steering mirrors or gimbal devices to control beam pointing directions \cite{milasevicius_2023_review,kaymak_2018_survey}. Meanwhile, small AO systems often rely on MEMS deformable mirrors for high-speed wavefront correction, though liquid crystal spatial light modulators are often used for low-speed systems \cite{zhang_2023_adaptive,akondi_2014_closed}. Due to the reliance on specialized components for each respective task, designing a communication or long-range imaging device that can perform both tasks with low SWaP is challenging. 

In this contribution, we demonstrate a series of correction algorithms and control systems that allow  the Texas Instruments Phase Light Modulator (PLM) \cite{bartlett_2021_recent} to act as a chip-scale standalone device for performing wavefront correction, acquisition, and target tracking, therefore significantly reducing the SWaP for the respective applications. We demonstrate that while using our methods, the PLM is capable of efficiently steering and tracking beams in real time, can correct real-time aberrations in an AO control loop, and can efficiently acquire and establish a communication link. Finally, we demonstrate novel acquisition techniques that we have exclusively tuned for spatial light modulator-based operation, to improve the speed of search and acquisition procedures.

\section{METHODS AND RESULTS}
\label{sec:bgmethods}
A PLM is a spatial light modulator, which directly changes the phase of incident light using an array of individually-actuated micromirrors, each moving in a piston motion. Each individual micromirror has a side length of 10.8 $\mu$m and can be individually addressed at high speeds. The 0.67" PLM model, for instance, hosts a 1358x800 micromirror array that can be controlled at speeds of up to 1.44 kHz. Subsequent models, such as the 0.98" PLM, have both higher resolutions, and speeds going beyond 5kHz \cite{bartlett_2021_recent}. These specifications have led to the exploration of the  PLM for many research tasks in the literature, such as beam steering \cite{deng_2022_diffraction} and holographic displays \cite{schiffers_2024_holochrome}. The specifications of the PLM lend themselves especially well towards FSoC-specific tasks. The high speed and resolution of the PLM could allow wavefront correction of scintillated beams up to high numbers of zernike polynomials, while the 4-bit depth boasts a steering error below 30 µrad for 1550 nm light\cite{nero_2025_simulating}, proving that PLM-based steering systems are feasible.

In previous publications, we have demonstrated open-loop AO correction with a PLM \cite{bass_2024_towards,chen_2025_adaptive}. However, the open-loop configuration used made system alignment difficult, and stopped the implementation and testing of search and acquisition routines.  Herein, we show a series of control algorithms that enable FSoC-specific tasks such as search and acquisition, tracking, and closed-loop AO correction to be performed with the PLM at high speeds and, due to the chip-scale nature of the PLM, with low SWaP of the final system.  In the following paragraphs, we will describe each respective control algorithm, as well as first demonstrations of the use of the PLM for all of these tasks.

\subsection{Experimental Apparatus}
\begin{figure} [h!]
    \centering
    \includegraphics[width=0.7 \linewidth]{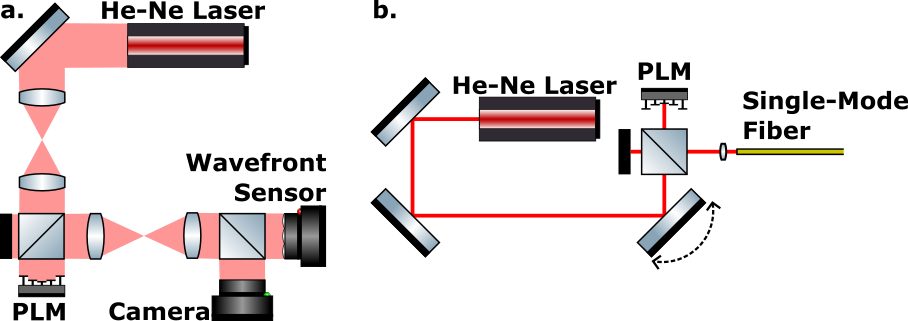}
    \caption{(a) Schematic of experimental system used to perform acquisition, tracking, and AO procedures described in Secs. \ref{subsec:closedloopao}, \ref{subsec:cameraao}, \ref{subsec:quadao}, and \ref{subsec:smartsearchandacquisition}. (b) Schematic of experimental system used to perform automatic fiber coupling as described in Sec. \ref{subsec:smfcoupling}}
    \label{fig:schematic}
\end{figure}

To demonstrate the use of PLM for each respective application, we used different variations of the basic experimental apparatus shown in Fig. \ref{fig:schematic}(a). The system consists of a two-lens telescope to relay an incident field to a PLM with unit magnification, followed by a second telescope to relay the modulated field onto both a camera and a wavefront sensor. In the following sections, we demonstrate four different experiments that gradually decrease the SWaP of the final system by modifying the control algorithm of the PLM. By using inputs from either the wavefront sensor (sec. \ref{subsec:closedloopao}) or the camera  (sec. \ref{subsec:cameraao}), we demonstrate the use of the PLM for adaptive optics, to perform wavefront correction up to five zernike polynomials. Additionally, we demonstrate tip-tilt correction using full-field camera feedback (sec. \ref{subsec:cameraao}), as well as feedback with an emulated quad-detector (sec. \ref{subsec:quadao}), and demonstrate fast beacon acquisition using an emulated single-pixel photodetector (sec. \ref{subsec:smartsearchandacquisition}).

Finally, to demonstrate the performance of beacon acquisition routines in practice, we set up a secondary experimental apparatus, shown in Fig. \ref{fig:schematic}(b), in which we demonstrated the use of PLM for search and acquisition by using one such routine to automatically couple light into a single-mode fiber (sec. \ref{subsec:smfcoupling}).

\subsection{Closed-loop adaptive optics correction}
\label{subsec:closedloopao}

\begin{figure}[t!]
    \centering
    \includegraphics[width=1\linewidth]{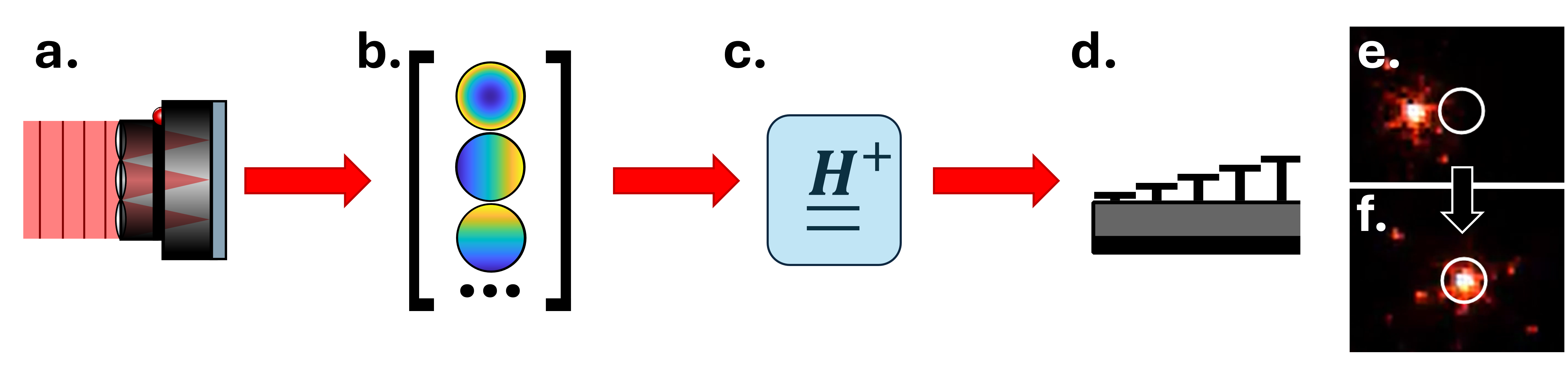}
    \caption{Procedure for AO correction with a PLM. (a) A wavefront sensor focuses an incident wavefront into a series of spots. (b) The spots are automatically fit to zernike polynomial amplitudes, representing the wavefront. (c) A series of two transmission matrices converts the zernike polynomial amplitudes to PLM mirror positions. (d) The PLM applies phase to remove wavefront aberrations. (e) A beam, as viewed by the camera depicted in Fig. \ref{fig:schematic}(a), before the tip and tilt are corrected to steer the beam to the center of a desired location, shown with a white circle. (f) The beam is moved to the desired location, after tip-tilt correction is performed using wavefront sensor feedback.}
    \label{fig:closedloopaoprocedure}
\end{figure}

By using feedback from the wavefront sensor shown in Fig. \ref{fig:schematic}(a), we demonstrate the correction of aberrations up to five zernike polynomials using a closed-loop control loop. To calculate the phase to apply to the PLM that corrects the aberrations sensed by the wavefront sensor, we implemented the transfer matrix-based correction procedure described in Eq. \ref{eq:plmPhase}.

\begin{equation}\label{eq:plmPhase}
\overrightarrow{\phi_{plm}} = K_p H_{plm} H_{transfer} \overrightarrow{z_{wfs}}
\end{equation}

In Eq. \ref{eq:plmPhase}, $\phi_{plm}$ is the pixel-by-pixel phase to apply to the PLM, $z_{wfs}$ is a vector containing measured zernike polynomial aberrations from the wavefront sensor, $H_{transfer}$ is a measured transfer matrix that converts wavefront sensor-plane zernike aberrations to PLM-plane zernike polynomial aberrations, $H_{plm}$ is a calculated transfer matrix which directly converts PLM-plane zernike polynomial aberrations to pixel-by-pixel phase levels, and $K_p$ is a proportional control term used to slow correction to prevent instability due to noise. Using this procedure, a measurement of a wavefront on the wavefront sensor can be quickly converted to a PLM phase correction pattern.

Using this procedure, we showed that the PLM could quickly decrease the wavefront error of a beam from multiple wavelengths to less than one wavelength by conjugating the phase created from the first five zernike modes (excluding piston). We found that our system would consistently lower the root mean square sum of first five zernike polynomial magnitudes after the beam was tilted and defocused by shifting and rotating upstream lenses and mirrors. Fig. \ref{fig:closedloopaoprocedure}(e-f) shows the observed results of simple tip-tilt correction, in which a beam is steered to a desired detector location using feedback from the wavefront sensor. The corresponding presentation video associated with this contribution additionally shows this correction in video form. We add that our method, in principle, also allows for the correction of higher-order zernike polynomials, but is currently limited by the SNR of the wavefront sensor, and aliasing on the PLM when wavefront error is high.

\subsection{Tip-tilt correction using camera feedback}
\label{subsec:cameraao}

\begin{figure}[hb!]
    \centering
    \includegraphics[width=1\linewidth]{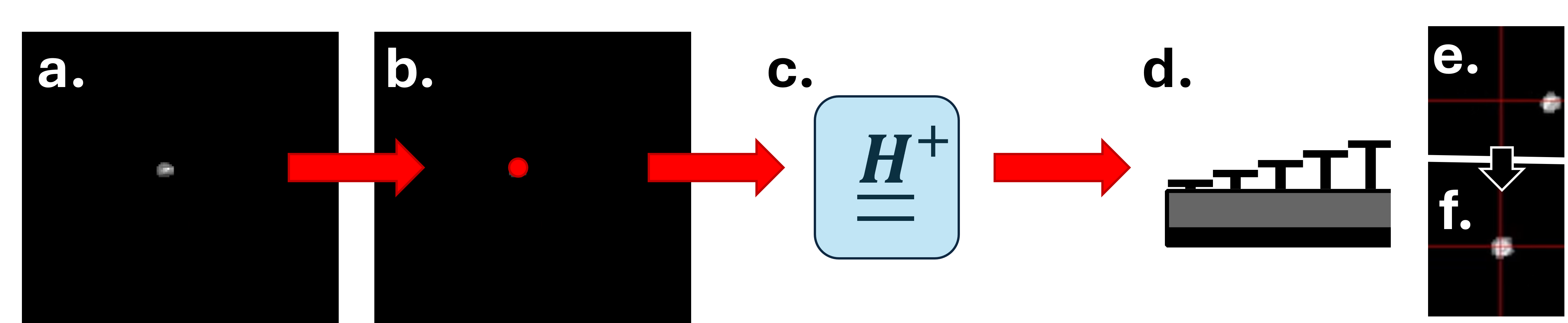}
    \caption{Procedure for PLM tip-tilt correction with camera feedback. (a) A camera captures an image of a focused beam. (b) The centroid position of the image is located. (c) The image centroid is converted to PLM mirror positions using two transmission matrices. (d) The PLM applies phase to correct beam pointing errors. (e) An captured beam on the camera, in an experiment, is initially not pointed towards a desired location marked with a red cross. (f) After camera feedback is applied, the PLM steers the beam to the center of the cross.}
    \label{fig:cameraaoprocedure}
\end{figure}

In specific scenarios (e.g., for small receive apertures) tip-tilt can be considered as the main form of aberration. AO correction of just tip and tilt can be considered a type of beam tracking, and thus can be implemented using feedback from lower-cost, lower SWaP devices than a wavefront sensor. Following this principle, we demonstrate tip-tilt correction using the PLM using feedback from a standard off-the-shelf camera.
To properly correct tip-tilt from camera information, we applied the correction procedure shown in Eq.~\ref{eq:plmPhase}, but replaced the vector $\overrightarrow{z_{wfs}}$ with a new vector $\overrightarrow{z_{cam}}$, containing offsets of the beam centroid from a target location in the image, and remeasured $H_{transfer}$ using this new input. This procedure is depicted in Fig. \ref{fig:cameraaoprocedure}(a-d). After applying the modified correction procedure, we demonstrated that the PLM could continuously steer a beam to the center of the camera, as shown in both Fig. \ref{fig:cameraaoprocedure}(e-f), and in video form within the corresponding presentation published alongside this proceeding. Using this correction procedure, a detector coupling apparatus could in principle be placed at a location corresponding to this focus position, which would allow the camera feedback to continuously allow light to be coupled into a photodetector, regardless of the incident beam tilt. 

\subsection{Tip-tilt correction using emulated quad detector feedback}
\label{subsec:quadao}

\begin{figure}[h!]
    \centering
    \includegraphics[width=1\linewidth]{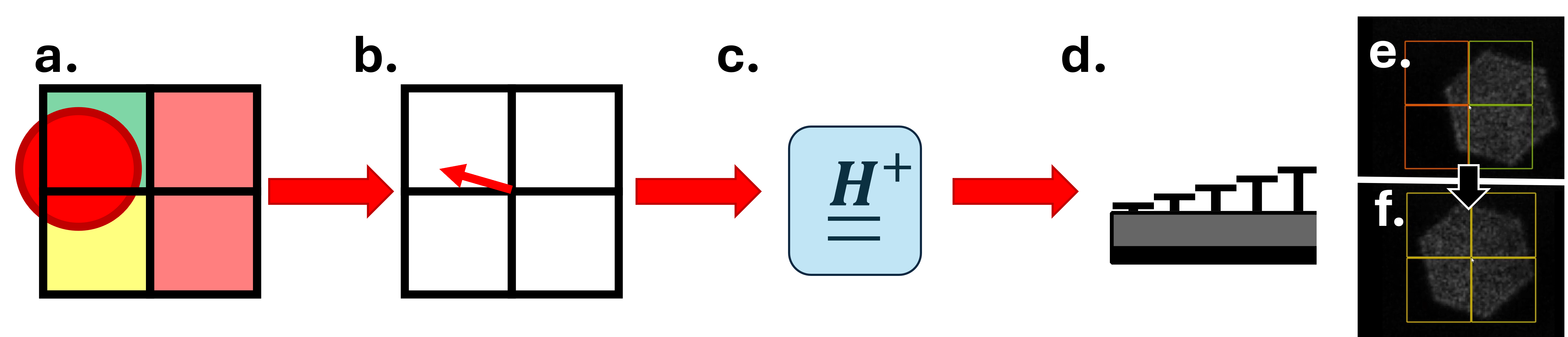}
    \caption{Procedure for PLM tip-tilt correction with quad detector feedback. (a) A quad detector samples the intensity of a beam in four different locations (b) The gradient vector of the quad detector pixels is calculated. (c) The image centroid is converted to PLM mirror positions using two transmission matrices. (d) The PLM applies phase to correct beam pointing errors. (e) In an experiment, an incident beam is offset from the center of the emulated quad-detector. (f) Using the feedback from the emulated quad detector, the PLM steers the beam back to the center of the detector.}
    \label{fig:quadaoprocedure}
\end{figure}

While cameras can provide feedback for beam tracking, quad detectors are more frequently used in FSoC scenarios. This is because they provide information about the offset centroid of a beam both faster and with lower SWaP compared to a camera. To demonstrate quad detector feedback without modifying our optical system, we emulated a quad detector using our camera by integrating the intensities over four larger camera regions, and using the four integrated intensity values as inputs of an assumed quad detector to our correction algorithm. Then, we demonstrated tip-tilt correction by adapting the same procedure that was used for camera-based correction, but replaced the vector $\overrightarrow{z_{cam}}$ with a new vector $\overrightarrow{z_{quad}}$ containing the horizontal and vertical intensity gradients of the emulated quad detector. Our approach is summarized in Fig. \ref{fig:quadaoprocedure} (a-d).

Using this approach, we found that the PLM could steer a beam to the center of the quad detector using only the computed gradients for feedback. An example of correction is shown in Fig. \ref{fig:quadaoprocedure} (e-f), and shown in video form within the corresponding presentation published alongside this proceeding.

\subsection{Search and acquisition for recieve-sided fiber coupling}
\label{subsec:smfcoupling}
The SWaP of a system can be reduced further by removing all external feedback-generating devices, and instead performing tip-tilt correction with the single-pixel photodetector used for communication. When doing so, the tip-tilt correction problem becomes a search and acquisition problem in transmit-side architectures, and a detector coupling problem in recieve-side architectures. One challenging variant of this is single-mode fiber coupling, which is required to build systems which make use of fiber-coupled photodetectors. 

Thus, to demonstrate that a PLM can speed up the fiber coupling process, we constructed the experimental apparatus shown in Fig. \ref{fig:schematic}(b), where a PLM modulates the phase of light before it is coupled into a single-mode fiber. Then, after establishing a link, we broke the coupling by rotating an upstream mirror, to establish a situation where the beam was close to being coupled, but still uncoupled. We demonstrated that we could reestablish the fiber coupling by scanning through tip-tilt combinations using the PLM until power began to enter the fiber. We additionally showed that we could improve the fiber coupling by performing a conical scan\cite{murphy_2009_conical} to progressively adjust the PLM coupling towards the direction of maximum power.

Finally, we demonstrated that existing coupling could be improved by correcting "static" aberrations such as spherical aberration or astigmatism, created by a slightly-misaligned optical system, using the PLM, to improve the fiber coupling further.

\subsection{Improved search and acquisition using adaptive beam defocusing}
\label{subsec:smartsearchandacquisition}

\begin{figure}[t!]
    \centering
    \includegraphics[width=1\linewidth]{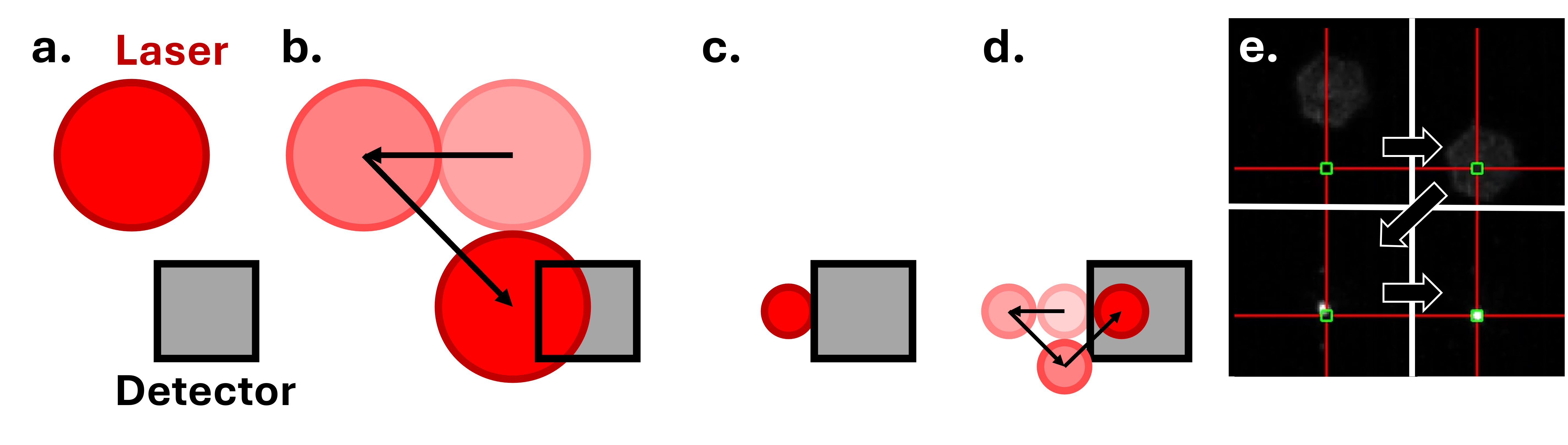}
    \caption{Procedure for search and acquisition using adaptive beam defocusing. (a) A beam starts at an unknown location relative to a detector, with a large set size. (b) By scanning the large beam, the detector is "roughly" found. (c) With the "rough" position known, the beam is set to a small size (d) Scanning with a small sized beam now quickly finds a precise detector location. (e) Acquisition procedure in a real experiment. The beam started large, before locating a simulated photodetector. Then, the beam size was lowered, and an additional search step, combined with a conical scanning procedure, refined the beam location to maximize the power on the simulated photodetector}
    \label{fig:smartsearchandacquisitionprocedure}
\end{figure}

The defocusing capabilities of the PLM can be used to increase the size of the beam in a search plane. Assuming there is enough power, this allows larger regions to be searched in a given search step. Using this principle, we developed an efficient acquisition method which adaptively increases the beam size when the detector position is unknown, and decreases the beam size as the detector position becomes known. While at a given beam size, a maximum-likelihood search method was used to choose angles to check, where we assumed that the likelihood of the receiver being at a location was inversely proportional to the angular distance from the initial pointing direction. The proposed search procedure is outlined further in Fig. \ref{fig:smartsearchandacquisitionprocedure}(a-d).

To test our method, we used the experimental system in Fig. \ref{fig:schematic}(a). To simulate a detector, we integrated over the central 5x5 pixel area on our camera, and used this as the only method of feedback for the search protocol. Using this procedure, we demonstrated efficient beam acquisition and tracking using the aforementioned adaptive beam size method. Our method located a simulated detector significantly faster than what a simple spiral scan or maximum-likelihood search could achieve without changing beam size. A demonstration of the proposed search and acquisition method, in a real experiment, is depicted in Fig. \ref{fig:smartsearchandacquisitionprocedure}(e), as well as shown in video form within the corresponding presentation published alongside this proceeding.

\section{DISCUSSION AND CONCLUSIONS}
\label{sec:conclusion}
The stated experiments demonstrate that our PLM-based framework is capable of performing wavefront correction, search, acquisition, and tracking procedures with minimal changes to system hardware, indicating that the PLM is versatile for multiple end applications such as FSoC. While our experiments in this contribution were done  for receive-side architectures, the methods used would also be applicable for transmit-side AO and acquisition/tracking systems. Future experiments will focus on the establishment of such transmit-side links, as well as the further refinement of smart acquisition protocols.

\section*{ACKNOWLEDGMENTS}       
This paper was prepared by the named authors in collaboration with the National Center for Manufacturing Sciences (NCMS). The research effort and this paper is/was sponsored by the funding from U.S. Department of Defense. The content of this report does not necessarily reflect the position or policy of NCMS or the Government; no official endorsement should be inferred.

This work was presented on August 7th, 2025 at the Unconventional Imaging, Sensing, and Adaptive Optics Conference at SPIE Optics and Photonics 2025, and will be published in SPIE Proceedings.

\section*{REFERENCES}
\bibliographystyle{ieeetr}
\bibliography{references}

\end{document}